\documentclass[aps,prl,twocolumn,amsmath,amssymb,superscriptaddress,nofootinbib]{revtex4-2}

\usepackage{graphicx}
\usepackage{tikz}
\usepackage{bm}
\newcommand{\dd}{\,\mathrm d}
\usepackage[colorlinks=true,allcolors=blue]{hyperref}

\newif\ifpending\pendingfalse

\begin{document}
\setcounter{secnumdepth}{1}

\title{A Finite Bid--Ask Spread from Replenishment Displaced from the Quote}

\author{Christopher Angstmann}
\email{c.angstmann@unsw.edu.au}
\affiliation{School of Mathematics and Statistics, University of New South
Wales, Sydney, NSW 2052, Australia}
\author{Derick Diana}
\email{derick.diana@alumni.uct.ac.za}
\affiliation{Department of Statistical Sciences, University of Cape Town,
Rondebosch 7701, Western Cape, South Africa}
\author{Tim Gebbie}
\email{tim.gebbie@uct.ac.za}
\affiliation{Department of Statistical Sciences, University of Cape Town,
Rondebosch 7701, Western Cape, South Africa}

\date{\today}

\begin{abstract}
We give a unified analytic account of a finite bid--ask spread in a two-field reaction--diffusion order book. The model retains separate bid and ask densities in operational time, with diffusion, cancellation, reaction and external order creation. % v1.3.0: qualify exact sector separation.
On the symmetric equal-coefficient branch, the imbalance field determines the reaction price, while the total standing density carries the spread geometry. A market-making source is introduced by withholding order placement over a finite interval around the reaction boundary, representing round-trip replenishment displaced from the quote. Under a separated, quasistationary and weak-overlap approximation, transport penetrates inward from the placement edges over the cancellation length, giving the quoted spread as the sub-threshold core of the placement-source-free interval. Market orders remove standing density at the observed quote, whereas replenishment is delivered outside it and must diffuse back across the penetration depth. We derive the resulting delayed and attenuated quote-level response in operational time. On that branch, finite spread and price impact arise from distinct response sectors, while their calendar-time appearance requires a separate observation clock.
\end{abstract}

\maketitle

\section{Introduction}
\label{sec:introduction}

In a single-front reaction--diffusion order book, the imbalance field $\phi=\rho_B-\rho_A$ defines one reaction price through its moving zero
\cite{bak1997price,toth2011anomalous,donier2015fully,benzaquen2018multi,bouchaud2018tqp}.
Here $\rho_B(x,u)$ and $\rho_A(x,u)$ are the standing bid and ask densities at log-price $x$ and operational time $u$.
A local bid--ask width can be assigned by introducing a positive imbalance threshold $\epsilon_\phi$.
Let $p_{\rm b}$ and $p_{\rm a}$ denote the implied bid and ask log-prices satisfying $\phi(p_{\rm b},u)=+\epsilon_\phi$ and $\phi(p_{\rm a},u)=-\epsilon_\phi$.
Linearisation about the reaction price gives, for $s=p_{\rm a}-p_{\rm b}$,
\begin{equation}
 s=\frac{2\epsilon_\phi}{|\partial_x\phi|},
 \qquad\text{so that}\qquad
 s\,|\partial_x\phi|=2\epsilon_\phi .
 \label{eq:slaving}
\end{equation}
This width is fixed by the same local imbalance gradient that sets the linear response to signed flow \cite{toth2011anomalous,donier2015fully}.
An imbalance-only model therefore provides an implied threshold spread but no independently varying placement scale.

Agents act on public information, private assessments and the order flow they observe.
The bid and ask density fields retain the aggregate effects of their order placement, revision, cancellation and execution \cite{bouchaud2018tqp}.
Diffusion represents aggregate order revision, carrying standing density into a depleted interval and tending to narrow the spread.
In the construction below, pending opposite-side replenishment widens the interval over which the market-making sector withholds new supply.
The observed spread is the sub-threshold part of that interval remaining after inward transport and cancellation.

The symmetric reaction term cancels when the bid and ask equations are subtracted.
With equal diffusion and cancellation coefficients, the resulting imbalance equation is linear for prescribed sources.
Donier \emph{et al.}~\cite{donier2015fully} use this reduction and neglect the intrinsic bid--ask width in their coarse-grained description.
Here we retain both side densities and their total $\rho=\rho_A+\rho_B$.
This preserves the standing-density geometry needed to distinguish source placement from the observed quotes.
On the symmetric equal-coefficient branch, source placement introduces an additional geometric scale while the imbalance field carries the reaction-price and impact response \cite{AngstmannGebbie2026JSTAT,AngstmannGebbie2026LMF}.

The model evolves in operational time $u$.
Calendar-time observations are obtained through $u=U(t)$, where $t$ denotes calendar time and $U$ is non-decreasing and right-continuous.
Completion laws, response kernels and lags are defined in $u$; operational lags have no universal calendar duration without specifying $U$ and its associated subordination \cite{AngstmannGebbie2026JSTAT}.

% \subsection{Discrete construction and continuum limit.}
On a uniform log-price lattice $x_j$ with spacing $\Delta x$, let $\rho_{i,j}^{n}$ denote the standing density on side $i\in\{A,B\}$ at operational time $u_n$.
With $\Delta u_n=u_{n+1}-u_n$ and $\Delta_x$ the discrete Laplacian, the update is
\begin{align}
 \rho_{i,j}^{n+1}-\rho_{i,j}^{n}
 =&\,\Delta u_n\bigl[D_i\Delta_x\rho_{i,j}^{n}
 -\nu_i\rho_{i,j}^{n}
 -\kappa\rho_{A,j}^{n}\rho_{B,j}^{n}\bigr] \nonumber \\
 &+ \Delta u_n s_{i,j}^{n}+M_{i,j}^{n},
 \label{eq:discreteRD}
\end{align}
Here $D_i$ is the diffusivity, $\nu_i$ the cancellation rate and $\kappa$ the symmetric reaction coefficient, which removes equal bid and ask volume.
The source rate $s_{i,j}^{n}$ combines external supply with paired replenishment on the opposite side, delayed by a causal completion law.
The impulses $M_{i,j}^{n}$ are density increments due to market-order flow, already integrated over the update interval.

The lattice spacing $\Delta x$ sets the numerical resolution and vanishes in the continuum limit.
On the separated branch considered below, the market-making sector maintains a finite placement standoff by withholding new supply from an interval.
Let $q_{\rm b}$ and $q_{\rm a}$ denote the bid and ask placement edges in log-price, so that the withheld interval is $(q_{\rm b},q_{\rm a})$ and its width is $\Sigma=q_{\rm a}-q_{\rm b}$.
This standoff is selected by the supply-placement law.
The source terms are where that sector enters.\footnote{This statement is branch-specific. Other source and boundary conditions can produce a finite spread without a placement-source-free interval \cite{smith2003statistical}; the claim here is only that those mechanisms do not select the standoff $\Sigma=q_{\rm a}-q_{\rm b}$ maintained by the market-making source on the branch studied below.}

The external supply has two complementary components, $s_i^{\rm ext}=s_i^{\rm lit}+s_i^{\rm lat}$, referenced to the placement edge $q_i$, with $q_B=q_{\rm b}$ and $q_A=q_{\rm a}$.
The latent source vanishes at that edge and approaches a finite reservoir rate in the far field; the lit source is non-zero at the edge and vanishes in the far field.
Both sources are supported outside $(q_{\rm b},q_{\rm a})$; standing density at the inward observed quotes $p_{\rm b}$ and $p_{\rm a}$ is supplied by diffusion from the placement edges.
For given placement, the stationary source problem determines the side density at its own edge, $\rho_i^q=\rho_i(q_i,u)$.
The external source profiles are given in Appendix~\ref{app:sourceprofiles}.

In Sec.~\ref{sec:continuum}, we give the continuum equations and the paired replenishment law linking round-trip completion to source placement.
In Sec.~\ref{sec:discussion} we derive the spread and quote response, then examine execution thresholds and the relation between spread and price impact.
Appendices~\ref{app:sourceprofiles}--\ref{app:capacity} give the external source profiles, distinguish density response from first passage, and derive finite-window quote capacity.

\section{Continuum limit.}
\label{sec:continuum}
In the continuum limit,
\begin{equation}
 \partial_u\rho_i
 =D_i\partial_{xx}\rho_i-\nu_i\rho_i
 -\kappa\rho_A\rho_B+s_i,
 \label{eq:continuumRD}
\end{equation}
with event impulses retained as measures when the forcing is not smooth. On the equal-coefficient branch $D_A=D_B=D$ and $\nu_A=\nu_B=\nu$, define
$\rho=\rho_A+\rho_B$ and $\phi=\rho_B-\rho_A$. Then the decomposition is exact:
\begin{align}
 \partial_u\phi&=D\partial_{xx}\phi-\nu\phi+(s_B-s_A),
 \label{eq:phi}\\
 \partial_u\rho&=D\partial_{xx}\rho-\nu\rho
 -\frac{\kappa}{2}(\rho^2-\phi^2)+(s_A+s_B).
 \label{eq:rho}
\end{align}
The symmetric reaction cancels from the imbalance equation and remains in the total-density equation.
Equal diffusion and cancellation coefficients give the closed linear imbalance equation \eqref{eq:phi} for prescribed sources.
Its zero $x_\phi$, where $\rho_A(x_\phi,u)=\rho_B(x_\phi,u)$, defines the reaction or latent log-price.
The quoted midpoint $m=(p_{\rm a}+p_{\rm b})/2$ is obtained from the observed quotes and coincides with $x_\phi$ on the symmetric branch.\footnote{With $D_A\neq D_B$ or $\nu_A\neq\nu_B$, the imbalance equation couples to total density and Eq.~\eqref{eq:phi} no longer applies. The symmetric reaction still cancels on subtraction, and the two-field construction remains valid.}

The placement profile of the paired source \eqref{eq:pairedsource} is supported outside $(q_{\rm b},q_{\rm a})$.
Let $\epsilon_B,\epsilon_A>0$ be the bid and ask standing-density thresholds used to resolve executable quotes.
These constitutive scales are supplied by the execution or observation rule and held fixed under numerical grid refinement.
The observed bid $p_{\rm b}$ is the highest log-price satisfying $\rho_B(p_{\rm b},u)=\epsilon_B$, and the observed ask $p_{\rm a}$ the lowest satisfying $\rho_A(p_{\rm a},u)=\epsilon_A$.
On the separated branch, these are inward level sets of the density solution.
The event-integrated impulses and causal completion law, together with replenishment initiated before the response-window origin, specify the forcing.

An execution at the ask opens a market-maker round trip whose closing leg is a purchase at the bid; an execution at the bid opens the reverse cycle.
We represent completion by paired opposite-side replenishment, with survival and execution probabilities included in the completion law and placement profile.
The replenishment current to side $i$ is $J_i^+(u)$, with placement profile $g_i(x,u)$.
Its paired source is
\begin{equation}
 s_i^{\rm mm}(x,u)=g_i(x,u)J_i^+(u).
 \label{eq:pairedsource}
\end{equation}
The profile is non-negative and normalised to unit integral over log-price, with bid support $x\leq q_{\rm b}$ and ask support $x\geq q_{\rm a}$.
For the displaced-kernel calculation below, $g_i(x,u)\simeq\delta(x-q_i(u))$ represents narrow placement at the corresponding edge.\footnote{For a distributed placement profile the quote response is the corresponding convolution, $K_i(\tau)=\int g_i(y)G_i^{|p_i-y|}(\tau)\dd y$; the single-distance formula is recovered when $g_i$ is narrow about $q_i$.}

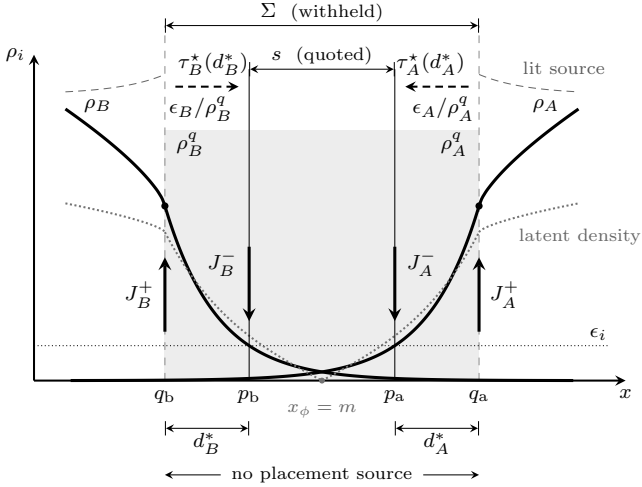
\begin{figure}[t]
\centering
\resizebox{\columnwidth}{!}{% Figure 1 -- displaced replenishment with latent density reference and lit source.
% v4 refined: latent density with source-free interior curvature, response
% groups moved above the shaded domain, extended guide lines.
% Black and white only: grey fill, no colour.
\begin{tikzpicture}[x=7.2mm,y=8.0mm,>=stealth,line join=round]

\def\qb{2}     \def\qa{8}      % placement quotes
\def\rq{3.0}   \def\eps{0.6}   % edge density and execution scale
\def\ds{1.609} % d* = ln(rho^q/epsilon)
\pgfmathsetmacro{\pb}{\qb+\ds}
\pgfmathsetmacro{\pa}{\qa-\ds}
\def\midp{5.0}
\def\ySigma{6.10}
\def\yspread{5.35}
\def\yresp{5.05}

% ---- source-free interval ------------------------------------------------
\fill[black!7] (\qb,0) rectangle (\qa,4.30);
\draw[black!45,dashed] (\qb,0) -- (\qb,\ySigma);
\draw[black!45,dashed] (\qa,0) -- (\qa,\ySigma);

% ---- axes ----------------------------------------------------------------
\draw[->,thick] (-0.5,0) -- (10.8,0) node[below] {\footnotesize $x$};
\draw[->,thick] (-0.5,0) -- (-0.5,5.60) node[left] {\footnotesize $\rho_i$};

% ---- execution scale -----------------------------------------------------
\draw[densely dotted] (-0.5,\eps) -- (10.5,\eps);
\node[anchor=south east,inner sep=1pt] at (10.5,\eps+0.04)
  {\footnotesize $\epsilon_i$};

% ---- style definitions ---------------------------------------------------
\tikzset{
  lit source/.style={thin,densely dashed,black!60},
  latent background/.style={line width=0.80pt,densely dotted,black!55}
}

% ---- standing densities --------------------------------------------------
\draw[very thick,domain=0.10:\qb,samples=40,smooth]
  plot (\x,{\rq+1.05*pow(\qb-\x,0.72)});
\draw[very thick,domain=\qb:9.8,samples=110,smooth]
  plot (\x,{\rq*exp(-(\x-\qb)/1)});
\draw[very thick,domain=\qa:9.90,samples=40,smooth]
  plot (\x,{\rq+1.05*pow(\x-\qa,0.72)});
\draw[very thick,domain=0.20:\qa,samples=110,smooth]
  plot (\x,{\rq*exp(-(\qa-\x)/1)});
\node[anchor=south,inner sep=2pt] at (0.72,4.52) {\footnotesize $\rho_B$};
\node[anchor=south,inner sep=2pt] at (9.28,4.52) {\footnotesize $\rho_A$};

% ---- latent density reference and lit source -----------------------------
% The dotted profile is a schematic latent-density reference. Inside the
% withheld region it follows a normalised source-free sinh profile so that it
% approaches x_phi=m linearly with finite slope and remains below the standing
% densities. Outside q_i it continues as a weaker background branch.
\def\latq{2.55}
\def\lout{2.55}
\def\lampa{0.32}
\def\ell{1.55}
\pgfmathsetmacro{\normsinh}{sinh((\midp-\qb)/\ell)}
% bid side
\draw[latent background,domain=0.10:\qb,samples=90,smooth]
  plot (\x,{\lout+\lampa*pow(\qb-\x,0.68)});
\draw[latent background,domain=\qb:\midp,samples=140,smooth]
  plot (\x,{\latq*sinh((\midp-\x)/\ell)/\normsinh});
% ask side, mirrored
\draw[latent background,domain=\qa:9.90,samples=90,smooth]
  plot (\x,{\lout+\lampa*pow(\x-\qa,0.68)});
\draw[latent background,domain=\midp:\qa,samples=140,smooth]
  plot (\x,{\latq*sinh((\x-\midp)/\ell)/\normsinh});

% Reaction price / midpoint marker on the symmetric branch.
\fill[black!55] (\midp,0) circle (1.15pt);
\node[anchor=north,inner sep=1.2pt,text=black!60] at (\midp,-0.35)
  {\scriptsize $x_\phi=m$};

% Lit external source: non-zero at q_i and decays outward, shown on a separate
% schematic scale and placed high enough not to collide with the rho labels.
\def\sbase{4.92}
\def\samp{0.34}
\draw[lit source,domain=0.10:\qb,samples=90,smooth]
  plot (\x,{\sbase+\samp*exp(-(\qb-\x)/0.90)});
\draw[lit source,domain=\qa:9.90,samples=90,smooth]
  plot (\x,{\sbase+\samp*exp(-(\x-\qa)/0.90)});

% Direct labels placed adjacent to the profiles.
\node[anchor=west,inner sep=1.0pt,text=black!65] at (8.78,5.38)
  {\scriptsize lit source};
\node[anchor=west,inner sep=1.0pt,text=black!60] at (8.70,2.48)
  {\scriptsize latent density};

% ---- edge densities ------------------------------------------------------
\fill (\qb,\rq) circle (1.4pt);
\fill (\qa,\rq) circle (1.4pt);
\node[anchor=south west,inner sep=1.5pt] at (\qb+0.14,\rq+0.78)
  {\footnotesize $\rho_B^{q}$};
\node[anchor=south east,inner sep=1.5pt] at (\qa-0.14,\rq+0.78)
  {\footnotesize $\rho_A^{q}$};

% ---- the four prices -----------------------------------------------------
\draw[thin] (\pb,0) -- (\pb,\yspread);
\draw[thin] (\pa,0) -- (\pa,\yspread);
\foreach \p/\lab in {\qb/{$q_{\rm b}$},\pb/{$p_{\rm b}$},\pa/{$p_{\rm a}$},\qa/{$q_{\rm a}$}}
  \node[anchor=north,inner sep=2pt] at (\p,-0.06) {\footnotesize \lab};

% ---- penetration depths --------------------------------------------------
\draw[|<->|] (\qb,-0.82) -- (\pb,-0.82)
  node[midway,below,inner sep=1.5pt] {\footnotesize $d_B^{*}$};
\draw[|<->|] (\pa,-0.82) -- (\qa,-0.82)
  node[midway,below,inner sep=1.5pt] {\footnotesize $d_A^{*}$};

% ---- standoff and observed spread ---------------------------------------
\draw[|<->|] (\qb,\ySigma) -- (\qa,\ySigma)
  node[midway,above,inner sep=1.5pt] {\footnotesize $\Sigma$ \ {\scriptsize (withheld)}};
\draw[|<->|] (\pb,\yspread) -- (\pa,\yspread)
  node[midway,above,inner sep=1.5pt] {\footnotesize $s$ \ {\scriptsize (quoted)}};

% ---- depletion at p, replenishment at q ---------------------------------
\draw[->,very thick] (\pa,2.30) -- (\pa,1.02);
\node[anchor=west,inner sep=2pt] at (\pa+0.12,2.02) {\footnotesize $J_A^-$};
\draw[->,very thick] (\pb,2.30) -- (\pb,1.02);
\node[anchor=east,inner sep=2pt] at (\pb-0.12,2.02) {\footnotesize $J_B^-$};

\draw[->,very thick] (\qa,0.84) -- (\qa,2.12);
\node[anchor=west,inner sep=2pt] at (\qa+0.10,1.48) {\footnotesize $J_A^+$};
\draw[->,thick,densely dashed] (\qa-0.20,\yresp) -- (\pa+0.20,\yresp);
\node[anchor=south,inner sep=1.5pt] at ({(\pa+\qa)/2-0.12},\yresp+0.12)
  {\footnotesize $\tau_A^\star(d_A^*)$};
\node[anchor=north,inner sep=1.5pt] at ({(\pa+\qa)/2+0.12},\yresp-0.10)
  {\footnotesize $\epsilon_A/\rho_A^{q}$};

\draw[->,very thick] (\qb,0.84) -- (\qb,2.12);
\node[anchor=east,inner sep=2pt] at (\qb-0.10,1.48) {\footnotesize $J_B^+$};
\draw[->,thick,densely dashed] (\qb+0.20,\yresp) -- (\pb-0.20,\yresp);
\node[anchor=south,inner sep=1.5pt] at ({(\qb+\pb)/2+0.12},\yresp+0.12)
  {\footnotesize $\tau_B^\star(d_B^*)$};
\node[anchor=north,inner sep=1.5pt] at ({(\qb+\pb)/2-0.12},\yresp-0.10)
  {\footnotesize $\epsilon_B/\rho_B^{q}$};

% ---- source-free interval label -----------------------------------------
% Width marker for the placement-free interval, placed below the d_i^* markers.
\def\ynoplace{-1.62}
\node[anchor=center,inner sep=1.0pt] (noplacement) at (5,\ynoplace)
  {\scriptsize no placement source};
\draw[<-] (\qb,\ynoplace) -- ([xshift=-3pt]noplacement.west);
\draw[->] ([xshift=3pt]noplacement.east) -- (\qa,\ynoplace);

\end{tikzpicture}}
\caption{Schematic of displaced replenishment on the symmetric separated branch. The solid curves are the standing bid and ask densities $\rho_B$ and $\rho_A$. Placement sources are withheld from $(q_{\rm b},q_{\rm a})$ (shaded), while standing density diffuses inward from the placement edges. The dotted profiles indicate a latent-book reference geometry and vanish at the reaction price $x_\phi=m$ on the symmetric branch; they are not additive components of the standing densities and are not the latent external source $s_i^{\rm lat}$. The dashed profiles show the lit external source $s_i^{\rm lit}$ outside the placement edges on an independent schematic scale. Under the quasistationary weak-overlap approximation, each standing density reaches its execution scale $\epsilon_i$ at $p_i$ after a penetration depth $d_i^*=\ell_{\nu,i}\ln(\rho_i^q/\epsilon_i)$, giving $s\simeq\Sigma-d_A^*-d_B^*$. Depletion $J_i^-$ acts at $p_i$, whereas replenishment $J_i^+$ is delivered at $q_i$; the lag-integrated quote-level density response is attenuated by $\epsilon_i/\rho_i^q$, while the displaced response kernel peaks at $\tau_i^\star(d_i^*)$ in operational time.}
\label{fig:displaced}
\end{figure}

Depletion $J_i^-\geq0$ removes standing volume at the observed quote $p_i$; replenishment $J_i^+\geq0$ supplies volume through $g_i$, concentrated at the placement edge $q_i$ under the narrow-placement approximation.
The response window begins at $u_0$, and $J_i^+=J_{i,<u_0}^++J_{i,[u_0,u)}^+$ includes replenishment initiated before $u_0$ and within the window.
Let $Q_i$ denote the volume still awaiting paired replenishment on side $i$, including prehistory.
The total pending volume $Q_\Sigma=Q_A+Q_B$ sets the placement standoff through $\Sigma=\Sigma_0+\chi_sQ_\Sigma$, where $\Sigma_0\geq0$ is the baseline standoff and $\chi_s\geq0$ measures its sensitivity to pending volume.
The signed difference $Q_\Delta=Q_A-Q_B$ shifts the placement centre.
The total can remain positive when the signed difference vanishes.
Figure~\ref{fig:displaced} illustrates the placement and observed quote geometry.

Retaining both $\rho$ and $\phi$ preserves the standing-density geometry.
On the separated branch considered here, the market-making supply rule sets the finite placement standoff.
Diffusion carries standing volume into a depleted interval over an operational time of order $s^2/\bar D$, where $\bar D=(D_A+D_B)/2$.
Reaction removes overlapping bid and ask volume, while cancellation removes standing volume wherever it is present.
These mechanisms determine the density response within the withheld interval $(q_{\rm b},q_{\rm a})$; its inward level sets define the observed quotes.

% =====================================================================
%======================================================================

Changes in execution rates, completion laws and prehistory alter the pending volumes and move the placement edges.
The profiles $g_i$ translate with those edges, so their contribution to the paired source \eqref{eq:pairedsource} follows from the placement law.
The pending modes are computed from executed volumes and completion weights before the density update; the observed quotes are then read from the resulting fields.
With immediate completion, $Q_\Sigma=Q_\Delta=0$ and $\Sigma=\Sigma_0$.
A standoff above that floor requires completion memory.
The pending round trips provide an inventory interpretation of the placement standoff \cite{stoll1978supply,amihud1980dealership,ho1981optimal,grossman1988liquidity,avellaneda2008high,gueant2013dealing}.
Adverse selection can also contribute to the spread \cite{glosten1985bid,huang1997components}; the present source closure retains the inventory channel.

\section{Discussion}
\label{sec:discussion}
The fields $\rho_i(x,u)$ represent standing volume per unit log-price; their integrals over a price interval give volume.
An interpretation in terms of the probability of individual order arrivals requires a further microscopic model.
The prescribed lit, latent and market-making sources vanish inside $(q_{\rm b},q_{\rm a})$, while standing volume can enter through aggregate order revision represented by diffusion.
In the stationary weak-overlap approximation, this inward transport balances cancellation; transient density also depends on the preceding state.
A displayed limit order submitted inside the spread raises the bid or lowers the ask in a discrete book.
Such a submission would enter this construction as an additional source event.

We consider bid and ask profiles whose overlap near each quote is small enough for cancellation to dominate reaction.
If diffusion and cancellation adjust the profiles faster than the placement edges and edge densities change, the leading balance inside the placement interval is $D_i\partial_{xx}\rho_i=\nu_i\rho_i$.
The resulting profile decays inward from each edge over the cancellation length $\ell_{\nu,i}=\sqrt{D_i/\nu_i}$.
For an edge density $\rho_i^{q}>\epsilon_i$, it reaches the execution threshold at depth $d_i^{*}=\ell_{\nu,i}\ln(\rho_i^{q}/\epsilon_i)$.
The observed quotes lie at $p_{\rm b}\simeq q_{\rm b}+d_B^{*}$ and $p_{\rm a}\simeq q_{\rm a}-d_A^{*}$, giving the leading quoted spread
\begin{equation}
 s\simeq\Sigma-d_A^{*}-d_B^{*}.
 \label{eq:spread}
\end{equation}
The quoted spread is the sub-threshold core remaining between the penetration layers (Fig.~\ref{fig:displaced}), with a positive separated spread requiring $\Sigma>d_A^{*}+d_B^{*}$.\footnote{Equation~\eqref{eq:spread} is an outer, separated-branch approximation. As $\Sigma\downarrow d_A^*+d_B^*$ the two penetration layers overlap and the approximation must be replaced by the coupled two-field problem; the formula should not be continued to negative $s$.}

The execution thresholds $\epsilon_i$ must be calibrated to the execution or observation rule; the density equations do not select them.
A possible calibration starts from the minimum standing volume that can be executed or resolved as a quote within a specified interval in log-price.
If the density varies little across that interval, dividing the volume by its width gives a density scale; the interval is held fixed under grid refinement.
For a fixed placement interval and exponential profile, increasing $\epsilon_i$ reduces penetration and widens the quoted spread, with $\partial s/\partial\ln\epsilon_i=\ell_{\nu,i}$.
At fixed $\Sigma,D_i,\nu_i$ and $\epsilon_i$, lower edge density $\rho_i^q$ also reduces penetration and widens the spread.
This relates a shallower book to a wider spread under the stated assumptions; changes in liquidity may also affect placement, diffusion and cancellation.

The exponential profile gives $|\partial_x\rho_i(p_i)|=\epsilon_i/\ell_{\nu,i}$ at the quote.
With $\epsilon_i$ fixed, a small density change $\delta\rho_i$ at the reference quote produces a displacement $\delta p_i\simeq-\delta\rho_i(p_i)/\partial_x\rho_i(p_i)$.
More bid density raises the bid; more ask density lowers the ask.
Where the density evolves smoothly and the quote slope is non-zero, differentiating the level-set condition gives $dp_i/du=-\partial_u\rho_i(p_i,u)/\partial_x\rho_i(p_i,u)$.
Quote motion follows the net density change after diffusion, cancellation and reaction.
In the stationary profile, transport balances the losses and the quotes do not move.
The factor $\ell_{\nu,i}/\epsilon_i$ sets the magnitude of the local quote response to a given density perturbation; the perturbation itself depends on quote location and applied flow.
The stationary source problem determines $\rho_i^q$ for the prescribed supply and placement; the external profiles are given in Appendix~\ref{app:sourceprofiles}.
The exponential approximation requires $\ell_{\nu,i}/s\ll1$ and weak reaction losses at the quote; a sufficient local condition is $\kappa\rho_{\rm opposite}(p_i)\ll\nu_i$, where $\rho_{\rm opposite}$ is the other side's density.

The source-free placement interval restricts the supply represented in this construction.
Other participants may submit orders within it.
At fixed placement, the stationary weak-overlap response to a bounded additional source $s_i^{\rm other}$ satisfies $\|\delta\rho_i\|_\infty\leq\|s_i^{\rm other}\|_\infty/\nu_i$.
Its effect on the observed quote is governed by the local density slope.
Additional supply can therefore be treated as a perturbation when the induced density and quote displacement are small on the execution and spread scales.

Depletion removes standing volume at the observed quote $p_i$, while the supply rule places replenishment at $q_i$.
For the linear diffusion--cancellation response, we hold these locations fixed and write $d=|q_i-p_i|>0$.
With $\tau>0$ denoting elapsed operational time since placement, the density response per unit supplied volume is
\begin{equation}
 G_i^{d}(\tau)=\frac{e^{-\nu_i\tau}}{\sqrt{4\pi D_i\tau}}\,
 \exp\!\left[-\frac{d^{2}}{4D_i\tau}\right],
 \label{eq:offset}
\end{equation}
Its maximum occurs at
\begin{equation}
 \tau_i^\star(d)
 =\frac{\sqrt{1+4\nu_i d^2/D_i}-1}{4\nu_i},
 \qquad
 \tau_i^\star(d)\longrightarrow\frac{d^2}{2D_i}
 \quad(\nu_i\downarrow0).
 \label{eq:peaklag}
\end{equation}
On the reference exponential profile $d=d_i^*$, so lowering $\epsilon_i$ at fixed $\rho_i^q,D_i,\nu_i$ increases the propagation distance and moves this maximum to a later operational lag.
This lag follows placement; the completion delay is already contained in $J_i^+$.
Determining the peak of the quoted-midpoint response requires the depletion and replenishment histories on both sides.
Calendar-time lags require the observation clock $U$.

Cancellation attenuates the density response as replenishment spreads from $q_i$ to $p_i$.
Holding these locations fixed and integrating over operational lag gives, at $d=d_i^*$,
\begin{equation}
 \frac{\int_0^{\infty}G_i^{d_i^{*}}(\tau)\dd\tau}
      {\int_0^{\infty}G_i^{0}(\tau)\dd\tau}
 =e^{-d_i^{*}/\ell_{\nu,i}}
 =\frac{\epsilon_i}{\rho_i^{q}}.
 \label{eq:suppression}
\end{equation}
For equal volumes on side $i$, the lag-integrated density response to displaced replenishment is smaller in magnitude than the response to local depletion, by the factor $\epsilon_i/\rho_i^q<1$.\footnote{The stationary exponential profile and the lag-integrated kernel have the same attenuation over $d_i^*$; its reciprocal is the ratio whose logarithm sets the penetration depth in Eq.~\eqref{eq:spread}.}
This statement concerns the perturbation response about the stationary book.
The kernel measures density at the quote; its relation to first-passage arrival is given in Appendix~\ref{app:firstpassage}.
For the exponential reference profile, dividing the integrated replenishment kernel by the quote slope magnitude gives $(\ell_{\nu,i}/\epsilon_i)\int_0^\infty G_i^{d_i^*}(\tau)\dd\tau=1/(2\nu_i\rho_i^q)$.
This coefficient measures the magnitude of quote displacement integrated over operational time per unit replenishment volume.
With $\rho_i^q,D_i,\nu_i$ fixed, the explicit $\epsilon_i$ dependence cancels in this coefficient, while the quote location, propagation lag and response shape change.
Midpoint impact per executed volume requires both quote responses and their depletion and replenishment currents.

% v1.3.0: state the frozen-geometry status of the displayed response.
Quote motion depends on the preceding depletion and replenishment currents.
With the quote and placement locations fixed, the linear density contribution from these currents is
\begin{align}
 \delta\rho_i(p_i,u)=&
 -\int_0^u G_i^0(u-v)J_i^-(v)\dd v \nonumber \\
 &~~~+\int_0^u G_i^{d_i^*}(u-v)J_i^+(v)\dd v,
 \label{eq:volterra}
\end{align}
Here $u=0$ denotes the response-window origin, and $J_i^+$ includes replenishment initiated before it.
The depletion and replenishment kernels differ because the currents act at different locations.
For a small perturbation, the quote displacement is $\delta p_i\simeq-\delta\rho_i(p_i,u)/\partial_x\rho_i(p_i)$.
The midpoint and spread responses are $\delta m=(\delta p_{\rm a}+\delta p_{\rm b})/2$ and $\delta s=\delta p_{\rm a}-\delta p_{\rm b}$.
For slowly moving quotes and placement edges, the kernel distances depend on the past source locations and the current quote.
The placement law and level-set conditions then give coupled Volterra equations for midpoint and spread.

We define the quote capacity over an operational-time window $\Delta u$ using a constant depletion current at the reference quote.
The capacity $\mathcal L_i(\Delta u)=J_i^-\Delta u/|\delta p_i(\Delta u)|$ is the volume removed per unit linearised quote displacement.
With the reference slope $\ell_i^0=|\partial_x\rho_i(p_i)|=\epsilon_i/\ell_{\nu,i}$, the short-window approximation for $\nu_i\Delta u\ll1$ is
\begin{equation}
\begin{aligned}
 \mathcal L_i(\Delta u)
 &=\ell_i^0\sqrt{\pi D_i\Delta u}
 \left[1+O(\nu_i\Delta u)\right]\\
 &=\epsilon_i\sqrt{\pi\nu_i\Delta u}
 \left[1+O(\nu_i\Delta u)\right].
\end{aligned}
 \label{eq:capacity}
\end{equation}
The exact finite-window expression is given in Appendix~\ref{app:capacity}.
The capacity varies as $\sqrt{\Delta u}$ for short windows and tends to zero as the window shrinks, so it does not define a non-zero instantaneous coefficient.
The full response retains the separate depletion and replenishment kernels in Eq.~\eqref{eq:volterra}; their different locations prevent an exact reduction to the local net current $J_i^- -J_i^+$.

The quoted midpoint $m$ is obtained from the bid and ask level sets, whereas $x_\phi$ is the zero of the imbalance field.
They coincide when the side profiles are reflections of one another about $x_\phi$ and the execution thresholds are equal.
A signed-flow perturbation can break this symmetry even with equal diffusion and cancellation coefficients.
For a fixed density solution, changing $\epsilon_i$ changes the quote level sets and can move $m$; $x_\phi$ is unchanged.
When depletion is applied at the observed quotes, changing the thresholds can also alter the subsequent density evolution.
The placement law introduces the additional scale $\Sigma$, while the midpoint and spread responses share the evolving density fields and flow histories.

The side densities and execution thresholds define the observed quotes; total density must be accompanied by the imbalance to distinguish bids from asks.
For prescribed sources, these level sets can define a spread without a separate market-making state.
The additional state used here is the volume awaiting opposite-side replenishment.
It sets the placement standoff through $\Sigma=\Sigma_0+\chi_s Q_\Sigma$, while the completion law determines when replenishment enters the book.
The supply geometry can therefore change with completion history while the execution thresholds remain fixed.
The market-making sector represents this aggregate replenishment dynamics.
Alternative source and placement laws can give different mechanisms for a finite spread.

\citet{wyart2008relation} constrain the spread--impact relation through a marginal-profitability condition on infinitesimal market-making and market-taking strategies.
We close the supply dynamics by linking the placement standoff to pending opposite-side replenishment, $\Sigma=\Sigma_0+\chi_s Q_\Sigma$.
The completion law determines when replenishment enters at the placement edges; diffusion and cancellation determine its inward density response.
The spread is read from the resulting penetration depths through Eq.~\eqref{eq:spread}, while quote motion follows the density changes at the execution thresholds.
The pending stock supplies the additional state linking execution and completion history to source placement and the subsequent quote response.

\section{Conclusion}
\label{sec:conclusion}

% v1.3.0: qualify the conclusion by the branch on which separation is exact.
We retain the side densities to distinguish source placement from the observed spread.
When reaction losses are small and the profiles adjust faster than source placement, diffusion and cancellation give $s\simeq\Sigma-d_A^{*}-d_B^{*}$ for separated quotes (Sec.~\ref{sec:discussion}, Eq.~\eqref{eq:spread}, and Fig.~\ref{fig:displaced}).
At fixed placement and with reaction neglected, the stationary density response acts as a spatial low-pass filter of order supply, with smoothing length $\ell_{\nu,i}$ in log-price.
Depletion acts at the observed quotes, whereas replenishment is delivered at the placement edges.
The displaced replenishment kernel peaks at a finite operational lag after placement, and cancellation attenuates its lag-integrated density response at the quote (Eqs.~\eqref{eq:offset}--\eqref{eq:suppression}; Appendix~\ref{app:firstpassage}).

Neither mechanism follows from the other.
A maker who sells at the ask completes the round trip by buying the same volume back at the bid.
We represent completion by paired opposite-side replenishment (Sec.~\ref{sec:continuum}).
This closes the cycle's inventory exposure, but need not balance the density response at the quotes: execution depletes volume at one quote, while replenishment is placed beyond the opposite quote and must diffuse inward, subject to cancellation (Sec.~\ref{sec:discussion}, Eqs.~\eqref{eq:offset}--\eqref{eq:suppression}).
The total volume awaiting completion, $Q_\Sigma$, sets the placement standoff and can remain positive even when signed inventory vanishes.
The spread follows from that standoff and the penetration depths in Eq.~\eqref{eq:spread}.
This introduces a dynamical link between round-trip completion, source placement and the spread beyond the single gradient that sets threshold width and linear impact in Eq.~\eqref{eq:slaving} (Sec.~\ref{sec:introduction}).

\section*{Reproducibility}
Code, saved numerical outputs and the computational supplement, with instructions for reproducing the numerical experiments, are available at \url{https://github.com/timgebbie/two-field-spread-reproducibility}.

\section*{AI Disclaimer}
% ---------------------------------------------------------------------
Claude (Opus 5) assisted with the initial draft iteration and bibliography checking from the concept paper. OpenAI ChatGPT (GPT-5.6 Sol) assisted with mathematical audit and revision suggestions, and helped with LaTeX/TikZ presentation. The authors reviewed all changes and are responsible for the results and final text. 
% ---------------------------------------------------------------------

\bibliographystyle{apsrev4-2}
\bibliography{CADDTG-SpreadLetter}

@article{AngstmannGebbie2026JSTAT,
  author        = {Angstmann, Christopher and Gebbie, Tim},
  title         = {Event-Time Order-Flow Memory, Operational-Time Impact, and
                   Subordinated Market Observables},
  journal       = {Journal of Statistical Mechanics: Theory and Experiment},
  year          = {2026},
  note          = {Accepted},
  eprint        = {2609.13715},
  archivePrefix = {arXiv},
  primaryClass  = {q-fin.TR},
  doi           = {10.48550/arXiv.2609.13715},
  url           = {https://arxiv.org/abs/2609.13715}
}

@misc{AngstmannGebbie2026LMF,
  author        = {Angstmann, Christopher and Gebbie, Tim},
  title         = {Revisiting Trade-sign Long-memory and Square-root Law price impact},
  year          = {2026},
  note          = {Extended version of arXiv:2609.13715},
  eprint        = {2606.16269},
  archivePrefix = {arXiv},
  primaryClass  = {q-fin.TR},
  doi           = {10.48550/arXiv.2606.16269},
  url           = {https://arxiv.org/abs/2606.16269}
}

@article{toth2011anomalous,
  title   = {Anomalous Price Impact and the Critical Nature of Liquidity in Financial Markets},
  author  = {T{\'o}th, Bence and Lemp{\'e}ri{\`e}re, Yves and Deremble, Cyril and de Lataillade, Joanne and Kockelkoren, Julien and Bouchaud, Jean-Philippe},
  journal = {Physical Review X},
  year    = {2011},
  volume  = {1},
  number  = {2},
  pages   = {021006},
  doi     = {10.1103/PhysRevX.1.021006}
}

@article{donier2015fully,
  title   = {A fully consistent, minimal model for non-linear market impact},
  author  = {Donier, Jonathan and Bonart, Julius and Mastromatteo, Iacopo and Bouchaud, Jean-Philippe},
  journal = {Quantitative Finance},
  year    = {2015},
  volume  = {15},
  number  = {7},
  pages   = {1109--1121},
  doi     = {10.1080/14697688.2015.1040056}
}

@article{benzaquen2018multi,
  title   = {Market impact with multi-timescale liquidity},
  author  = {Benzaquen, Michael and Bouchaud, Jean-Philippe},
  journal = {Quantitative Finance},
  year    = {2018},
  volume  = {18},
  number  = {11},
  pages   = {1781--1790},
  doi     = {10.1080/14697688.2018.1444279}
}

@book{bouchaud2018tqp,
  title     = {Trades, Quotes and Prices: Financial Markets Under the Microscope},
  author    = {Bouchaud, Jean-Philippe and Bonart, Julius and Donier, Jonathan and Gould, Martin D.},
  publisher = {Cambridge University Press},
  address   = {Cambridge},
  year      = {2018},
  doi       = {10.1017/9781316659335}
}

@article{bak1997price,
  title   = {Price variations in a stock market with many agents},
  author  = {Bak, Per and Paczuski, Maya and Shubik, Martin},
  journal = {Physica A: Statistical Mechanics and its Applications},
  year    = {1997},
  volume  = {246},
  number  = {3--4},
  pages   = {430--453},
  doi     = {10.1016/S0378-4371(97)00401-9}
}

@article{smith2003statistical,
  title   = {Statistical theory of the continuous double auction},
  author  = {Smith, Eric and Farmer, J. Doyne and Gillemot, L{\'a}szl{\'o} and Krishnamurthy, Supriya},
  journal = {Quantitative Finance},
  year    = {2003},
  volume  = {3},
  number  = {6},
  pages   = {481--514},
  doi     = {10.1088/1469-7688/3/6/307}
}

@article{wyart2008relation,
  title   = {Relation between bid--ask spread, impact and volatility in order-driven markets},
  author  = {Wyart, Matthieu and Bouchaud, Jean-Philippe and Kockelkoren, Julien and Potters, Marc and Vettorazzo, Michele},
  journal = {Quantitative Finance},
  year    = {2008},
  volume  = {8},
  number  = {1},
  pages   = {41--57},
  doi     = {10.1080/14697680701344515}
}

@article{stoll1978supply,
  title   = {The Supply of Dealer Services in Securities Markets},
  author  = {Stoll, Hans R.},
  journal = {The Journal of Finance},
  year    = {1978},
  volume  = {33},
  number  = {4},
  pages   = {1133--1151},
  doi     = {10.1111/j.1540-6261.1978.tb02053.x}
}

@article{amihud1980dealership,
  title   = {Dealership market: Market-making with inventory},
  author  = {Amihud, Yakov and Mendelson, Haim},
  journal = {Journal of Financial Economics},
  year    = {1980},
  volume  = {8},
  number  = {1},
  pages   = {31--53},
  doi     = {10.1016/0304-405X(80)90020-3}
}

@article{ho1981optimal,
  title   = {Optimal dealer pricing under transactions and return uncertainty},
  author  = {Ho, Thomas and Stoll, Hans R.},
  journal = {Journal of Financial Economics},
  year    = {1981},
  volume  = {9},
  number  = {1},
  pages   = {47--73},
  doi     = {10.1016/0304-405X(81)90020-9}
}

@article{glosten1985bid,
  title   = {Bid, ask and transaction prices in a specialist market with heterogeneously informed traders},
  author  = {Glosten, Lawrence R. and Milgrom, Paul R.},
  journal = {Journal of Financial Economics},
  year    = {1985},
  volume  = {14},
  number  = {1},
  pages   = {71--100},
  doi     = {10.1016/0304-405X(85)90044-3}
}

@article{grossman1988liquidity,
  title   = {Liquidity and Market Structure},
  author  = {Grossman, Sanford J. and Miller, Merton H.},
  journal = {The Journal of Finance},
  year    = {1988},
  volume  = {43},
  number  = {3},
  pages   = {617--633},
  doi     = {10.1111/j.1540-6261.1988.tb04594.x}
}

@article{huang1997components,
  title   = {The Components of the Bid-Ask Spread: A General Approach},
  author  = {Huang, Roger D. and Stoll, Hans R.},
  journal = {The Review of Financial Studies},
  year    = {1997},
  volume  = {10},
  number  = {4},
  pages   = {995--1034},
  doi     = {10.1093/rfs/10.4.995}
}

@article{avellaneda2008high,
  title   = {High-frequency trading in a limit order book},
  author  = {Avellaneda, Marco and Stoikov, Sasha},
  journal = {Quantitative Finance},
  year    = {2008},
  volume  = {8},
  number  = {3},
  pages   = {217--224},
  doi     = {10.1080/14697680701381228}
}

@article{gueant2013dealing,
  title   = {Dealing with the inventory risk: a solution to the market making problem},
  author  = {Gu{\'e}ant, Olivier and Lehalle, Charles-Albert and Fernandez-Tapia, Joaquin},
  journal = {Mathematics and Financial Economics},
  year    = {2013},
  volume  = {7},
  number  = {4},
  pages   = {477--507},
  doi     = {10.1007/s11579-012-0087-0}
}

\appendix

\section{Lit and latent source matching}
\label{app:sourceprofiles}
Let $d\geq0$ denote distance outward from the placement edge on either side. The continuum source profiles may be written as
\begin{align}
 s_i^{\rm lit}(d)&=\Lambda_i^{\rm lit}e^{-d/\ell_i^{\rm lit}},
 \label{eq:litsourceprofile}\\
 s_i^{\rm lat}(d)&=\Lambda_i^{\rm lat}
 \left(1-e^{-d/\ell_i^{\rm lat}}\right).
 \label{eq:latsourceprofile}
\end{align}
They have complementary limits at the placement edge,
\begin{equation}
 s_i^{\rm lit}(0)=\Lambda_i^{\rm lit},
 \qquad
 s_i^{\rm lat}(0)=0,
 \label{eq:sourceedge}
\end{equation}
and in the far field,
\begin{equation}
 s_i^{\rm lit}(d)\longrightarrow0,
 \qquad
 s_i^{\rm lat}(d)\longrightarrow\Lambda_i^{\rm lat}
 \quad(d\to\infty).
 \label{eq:sourcefarfield}
\end{equation}
Thus the combined external supply $s_i^{\rm ext}=s_i^{\rm lit}+s_i^{\rm lat}$ contains both a finite far-field reservoir and a non-zero near-edge supply. The distinction is one of source support and limiting behaviour, not of density support. Both external source profiles terminate at the placement edge $q_i$; because $q_{\rm b}<p_{\rm b}\leq p_{\rm a}<q_{\rm a}$, neither source is active at the observed level sets. Standing density at $p_i$ is generated by transport inward from the placement edge. The edge density $\rho_i^q$ used in Eq.~\eqref{eq:spread} is therefore an output of the stationary problem with the prescribed supply and placement, not an independent source amplitude.

\section{Density response and first passage}
\label{app:firstpassage}
\setcounter{equation}{0}
Equation~\eqref{eq:offset} is the killed-diffusion point-density Green function. It gives the density response observed at the quote after a perturbation is placed a distance $d$ away; it is not itself a first-passage density. If the question is instead the distribution of first arrival at the quote, the killed first-passage density is
\begin{equation}
 h_i^d(\tau)=\frac{d}{\sqrt{4\pi D_i\tau^3}}
 \exp\!\left[-\nu_i\tau-\frac{d^2}{4D_i\tau}\right].
 \label{eq:firstpassage}
\end{equation}
Its integrated survival-to-arrival probability is
\begin{equation}
 \int_0^\infty h_i^d(\tau)\dd\tau=e^{-d/\ell_{\nu,i}},
 \label{eq:firstpassageint}
\end{equation}
so at $d=d_i^*$ it gives the same exponential attenuation $\epsilon_i/\rho_i^q$ as the lag-integrated density response. The lag of the maximum is different:
\begin{equation}
 \tau_{i,{\rm FP}}^\star(d)
 =\frac{\sqrt{9+4\nu_i d^2/D_i}-3}{4\nu_i}
 \longrightarrow \frac{d^2}{6D_i}
 \quad(\nu_i\downarrow0).
 \label{eq:firstpassagepeak}
\end{equation}
The Letter uses the density kernel because the moving level set responds to standing density. Equations~\eqref{eq:firstpassage}--\eqref{eq:firstpassagepeak} show that the same integrated attenuation may also be read as a survival-to-arrival probability when a literal first-passage interpretation is required, but the two lag distributions should not be identified.

\section{Finite-window capacity}
\label{app:capacity}
\setcounter{equation}{0}
For a constant same-location perturbation over a window $T=\Delta u$, the killed heat kernel gives
\begin{equation}
 I_i(T)=\int_0^T G_i^0(\tau)\dd\tau
 =\frac{1}{2\sqrt{D_i\nu_i}}\,
 \operatorname{erf}\!\sqrt{\nu_iT}.
 \label{eq:kernelint}
\end{equation}
If the level-set displacement is locally linearised using the slope magnitude $\ell_i^0=|\partial_x\rho_i|$, the corresponding finite-window capacity is
\begin{equation}
 \mathcal L_i(T)
 =\frac{\ell_i^0 T}{I_i(T)}
 =\frac{2\ell_i^0\sqrt{D_i\nu_i}\,T}
 {\operatorname{erf}\!\sqrt{\nu_iT}}.
 \label{eq:capacityexact}
\end{equation}
For $\nu_iT\ll1$,
\begin{equation}
 \mathcal L_i(T)
 =\ell_i^0\sqrt{\pi D_iT}
 \left[1+\frac{\nu_iT}{3}+O((\nu_iT)^2)\right].
 \label{eq:capacityexpansion}
\end{equation}
Thus Eq.~\eqref{eq:capacity} is the short-window asymptotic, not an exact finite-window identity. More generally, when depletion acts at $p_i$ and replenishment is placed at $q_i$, Eq.~\eqref{eq:volterra} is the primitive linear response; a single local forcing $F_i=J_i^- -J_i^+$ is not exact because the two currents are filtered by different kernels.

\end{document}